\input{aipcheck}

\documentclass[
    ,final            % use final for the camera ready runs
  ]
  {aipproc}

\layoutstyle{8x11double}

\begin{document}

\title{Red Giants observed with CoRoT\thanks{The CoRoT space mission which was developed and is operated by the French space agency CNES, with participation of ESA's RSSD and Science Programmes, Austria, Belgium, Brazil, Germany, and Spain.}}


\author{S. Hekker}{
  address={University of Birmingham, School of Physics and Astronomy, Edgbaston, Birmingham B15 2TT, UK}
   ,altaddress={Instituut voor Sterrenkunde, K.U. Leuven, Celestijnenlaan 200 D, 3001 Leuven, Belgium}
}

\author{J. De Ridder}{
  address={Instituut voor Sterrenkunde, K.U. Leuven, Celestijnenlaan 200 D, 3001 Leuven, Belgium}
}

\author{F. Baudin}{
 address={Institute d'Astrophysique Spatiale, UMR 8617, Universit\'e Paris XI, B\^atiment 121, 91405 Orsay Cedex, France}
}

\author{C. Barban}{
 address={LESIA, UMR 8109, Universit\'e Pierre et Marie Curie, Universit\'e Denis Diderot, Observatoire de Paris, 92195 Meudon Cedex, France}
}

\author{F. Carrier}{
  address={Instituut voor Sterrenkunde, K.U. Leuven, Celestijnenlaan 200 D, 3001 Leuven, Belgium}
}

\author{A.P. Hatzes}{
  address={Th\"uringer Landessternwarte, D-07778 Tautenburg, Germany}
}

\author{T. Kallinger}{
 address={Department of Physics and Astronomy, University of British Columbia, 6224 Agricultural Road, Vancouver, BC V6T 1Z1, Canada},
  altaddress={Institute for Astronomy, University of Vienna, T\"urkenschanzstrasse 17 A-1180 Vienna, Austria}
}

\author{W.W. Weiss}{
  address={Institute for Astronomy, University of Vienna, T\"urkenschanzstrasse 17 A-1180 Vienna, Austria}
}

\begin{abstract}
Observations of red (G-K)-giant stars with the CoRoT satellite provide unprecedented information on the stochastically excited oscillations in these stars. The long time series of nearly uninterrupted high-cadence and high-precision photometry revealed the presence of non-radial modes with long lifetimes, which opens the possibility to perform asteroseismology on these stars. Also, the large number of red giants, for which solar-like oscillations are now observed, allows for a more statistical investigation of the characteristics of solar-like oscillations in red giants.
\end{abstract}

\maketitle

%%%%%%%%%%%%%%%%%%%%%%%%%%%%%%%%%%%%%%%%%%%%
%% MAINMATTER
%%%%%%%%%%%%%%%%%%%%%%%%%%%%%%%%%%%%%%%%%%%%

\section{Introduction}
All stars with masses between roughly 0.5 and 10 solar masses evolve through a red-giant phase. 
%In this phase the star exhausted the hydrogen fuel in its core, which reduces the nuclear reaction rate, and the inner core contracts under its own gravity. This contraction heats a hydrogen shell outside the core, initiating hydrogen shell burning with an increased nuclear reaction rate due to the increased temperature, which results in an enhanced luminosity. Due to this enhanced luminosity the outer atmosphere expands greatly, which results in a lower surface temperature. Due to these changes in luminosity and temperature, 
Stars in this evolutionary phase follow a relatively narrow track through the H-R diagram, i.e., during the ascent and descent of the giant branch and their time on the horizontal branch, while their internal structures are presumably different as a function of e.g., mass, metallicity and primordial rotational velocity. Because so many stars become red giants, more information on the internal structure of these stars could improve our knowledge of stellar evolution considerably. Asteroseismology through stochastically excited oscillations can provide us with information about the internal structure of these stars and would therefore be the ideal way to acquire further knowledge of red-giant evolution.

Acoustic noise in the resonant cavity of stars with turbulent outer layers, such as red giants, drive intrinsically stable p-mode pulsations, i.e., stochastically excited and damped (solar-like) oscillations. Indeed, observational evidence of such oscillations in red giants was presented by \cite{frandsen2002}, followed by spectroscopic ground-based observations \cite{deridder2006,barban2004} and space-based photometry, e.g. \cite{barban2007,tarrant2008,stello2008,kallinger2008}. 

To use the observed solar-like oscillations for asteroseismology, frequency determination and mode identification are needed.  Up to now, the number of red giants with observed solar-like oscillations and the time-spans of these observations have been limited, leading to different interpretations of the modes in terms of mode-lifetimes and the presence of non-radial oscillations. A detailed discussion on these different interpretations is given by \cite{deridder2009}. The lack of accurate frequencies and mode identifications hampered the application of asteroseismology for red-giant stars.

The CoRoT (Convection, Rotation, and planetary Transits) satellite continuously observed $\sim$ 11\,400 stars for about 150 days during its first long run in the exo-planet field. Among these stars are many red giants and these long, nearly uninterrupted time series of high-cadence (32 seconds or 512 seconds)  and high-precision photometric data allow to determine accurate frequencies and to perform mode identification, as presented by \cite{deridder2009}. These results pave the way to investigations of the internal structure of red-giant stars by means of asteroseismology.

In these proceedings we will discuss some challenges in the field of red-giant asteroseismology and expand on the results from CoRoT presented by \cite{deridder2009} and the statistical investigation of the characteristics of solar-like oscillation in red giants \cite{hekker2009}. We finish with a discussion about future prospects of red-giant asteroseismology.

\section{Challenges}
The field of red-giant asteroseismology is still in its infancy, as this is both observationally as well as theoretically a challenging field. To infer the internal structure of the stars we need to determine a number of parameters, such as the frequency of maximum oscillation power, which provides information on the stellar radius \cite{kjeldsen1995} which increases with age. In cases of low-order high-degree modes, the asymptotic approximation \cite{tassoul1980} might be applied, and the large separation between modes with the same degree and consecutive orders can be used as a proxy for the average density of the stars. When knowing the individual frequencies and their degree an exact model for the star can be computed. Furthermore, the solar-like oscillations are stochastically excited and have finite mode lifetimes, i.e., the modes are excited and damped, with a rate $e^{-\eta t}$, with $\eta$ the damping rate, and re-excited before the oscillation mode is damped out. Due to this effect both the amplitude and phase of the oscillation are time dependent and the frequency peaks in the power spectrum have Lorentzian shapes with a width inversely proportional to the mode lifetime ($\tau$ $\sim$ $\eta^{-1}$). Therefore, the mode lifetime provides information on the excitation and damping processes, and on the turbulence parameters in the vicinity of the excitation of the modes.

So far, both theoretical and observational problems have prevented us from determining the oscillation parameters we just described. To name a few theoretical problems: the computation of convection, which is important for the driving and damping of the oscillations, is still difficult. The often used the classical mixing-length parametrisation \cite{bohmvitense1958} is computationally attractive, but does not provide a physical description of the convection processes within a star. Furthermore, the numerical stability of some evolution codes is decreased when modelling a helium flash scenario at the tip of the giant branch. This can be solved by computationally expensive algorithms or by circumventing the numerical problems by restarting the computation at the horizontal branch, see e.g. \cite{mazumdar2009}. Another numerical difficulty originates from the very large number of nodes in the eigenfunctions in the g-mode cavity. For state-of-the-art models of red-giant stars, we refer to \cite{dupret2009}.

The observational challenges are caused by the life-times of the oscillation modes, their amplitudes and their frequency range. First of all, the finite mode lifetimes of stochastically driven and damped oscillations are different for modes of different degrees, depending on their mode inertia. Higher mode inertia for non-radial modes lead to longer lifetimes, generally of the order of tens of days, but they can be of the order of hundreds of days or more, see e.g. \cite{dupret2009}. If the duration of the observations is shorter than these lifetimes, the modes are unresolved and have lower heights in the power spectrum. In order to resolve, and thus observe, the non-radial modes, we need time series of at least several tens of days, but preferably of hundreds of days.

A second observational challenge are the low amplitudes of the solar-like oscillations. Amplitudes depend on the height and width of the frequency peaks and scale with luminosity. For these intrinsically bright stars the heights of the frequency peaks are of the order of tens to hundreds of ppm in photometry or m\,s$^{-1}$ in radial velocity. This is indeed large compared to the solar values of a few ppm and cm\,s$^{-1}$, but still observationally only feasible for the last decade or so. 

Thirdly, the large radii of red-giant stars adjust the pulsation periods from minutes, for the Sun, to hours. This complicates ground-based single-site efforts considerably, and calls for either multi-site campaigns or observations from space. Furthermore, like for the Sun, turbulent motions in the convective envelope cause granulation. The granulation time scales depend on the size of the granules, i.e., the depth of the convective zone, and their velocities, and are generally of the order of days for red giants. For the most evolved and luminous stars, which have the largest radii, the granulation time scales are not resolved, which imposes observational difficulties to disentangle these effects.

The realisation of the CoRoT satellite, and more recently the Kepler satellite, are very important for tackling the observational challenges of observing solar-like oscillations in red giants. These satellites observe the same field for 150 days and 3.5 years, respectively, with high-precision photometry. This means that non-radial modes can be resolved and have detectable heights in the power spectrum.

\section{CoRoT results}

\begin{figure}
\begin{minipage}{\linewidth}
\centering
\includegraphics[width=\linewidth]{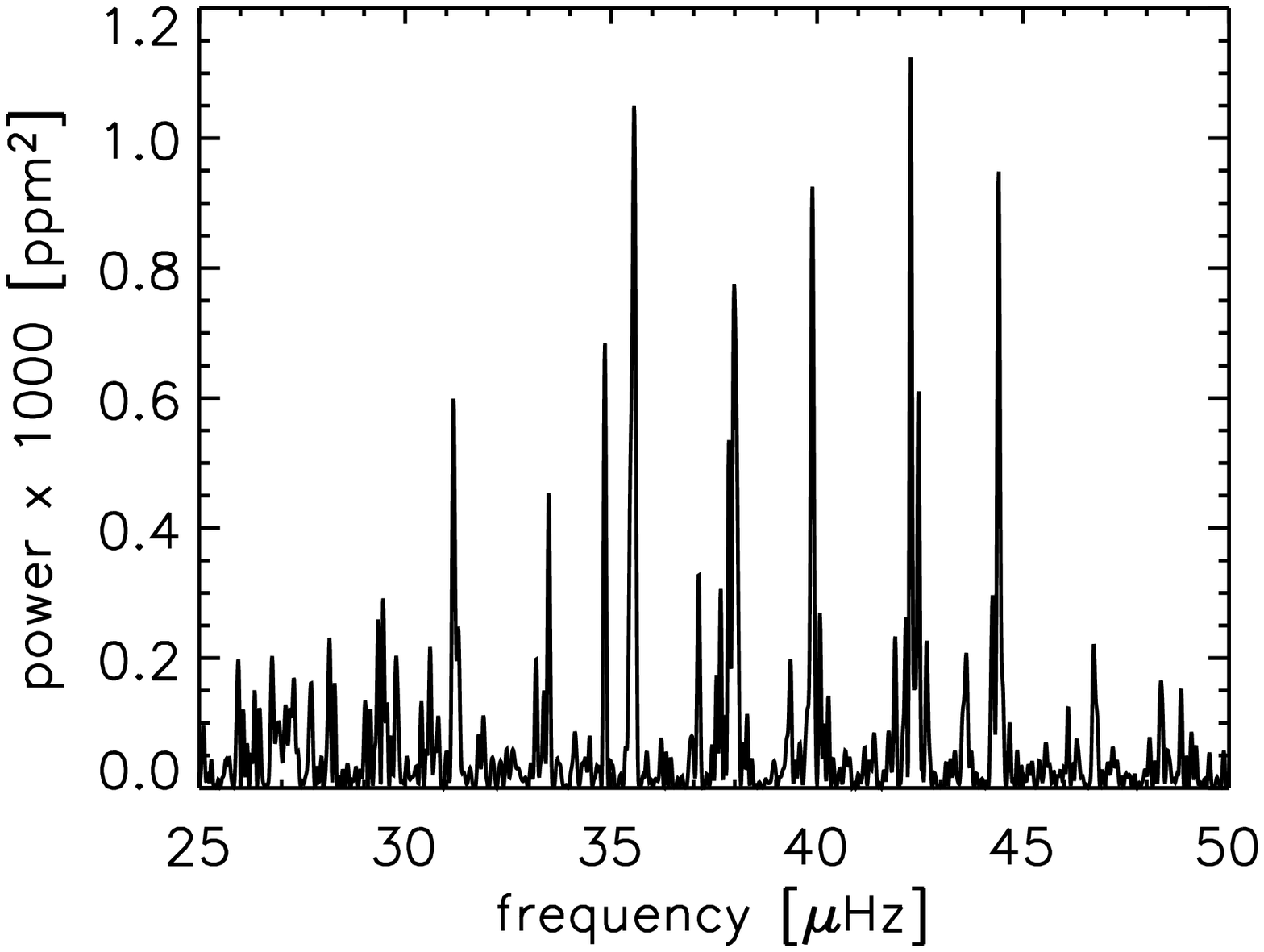}
\end{minipage}
\hfill
\begin{minipage}{\linewidth}
\centering
\includegraphics[width=\linewidth]{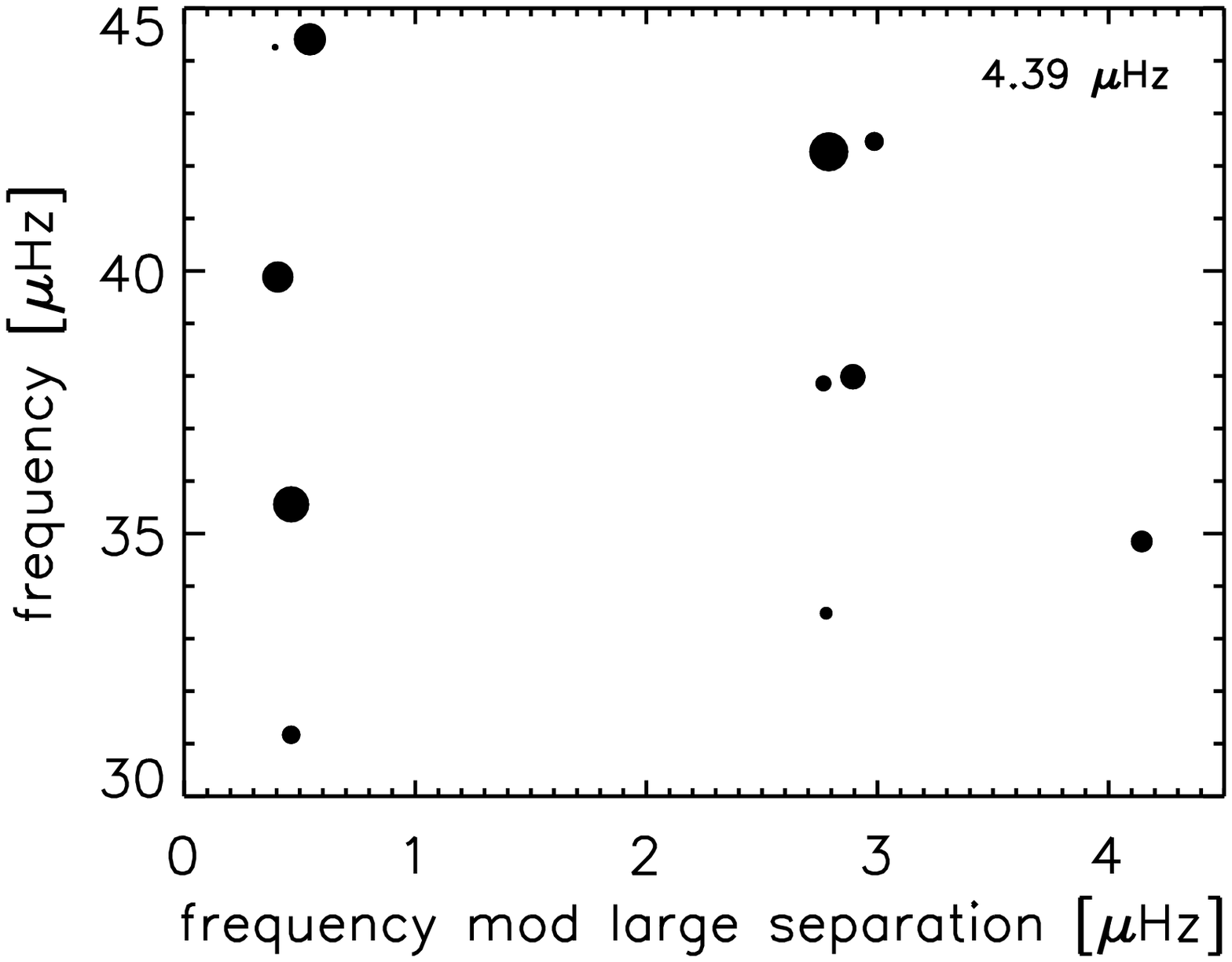}
\end{minipage}
\caption{Left: power spectrum of the CoRoT star 100483847. Right: \'echelle diagram of 100483847 with a large separation of 4.39 $\mu$Hz.}
\label{nonrad}
\end{figure}

\begin{figure}
\begin{minipage}{\linewidth}
\centering
\includegraphics[width=\linewidth]{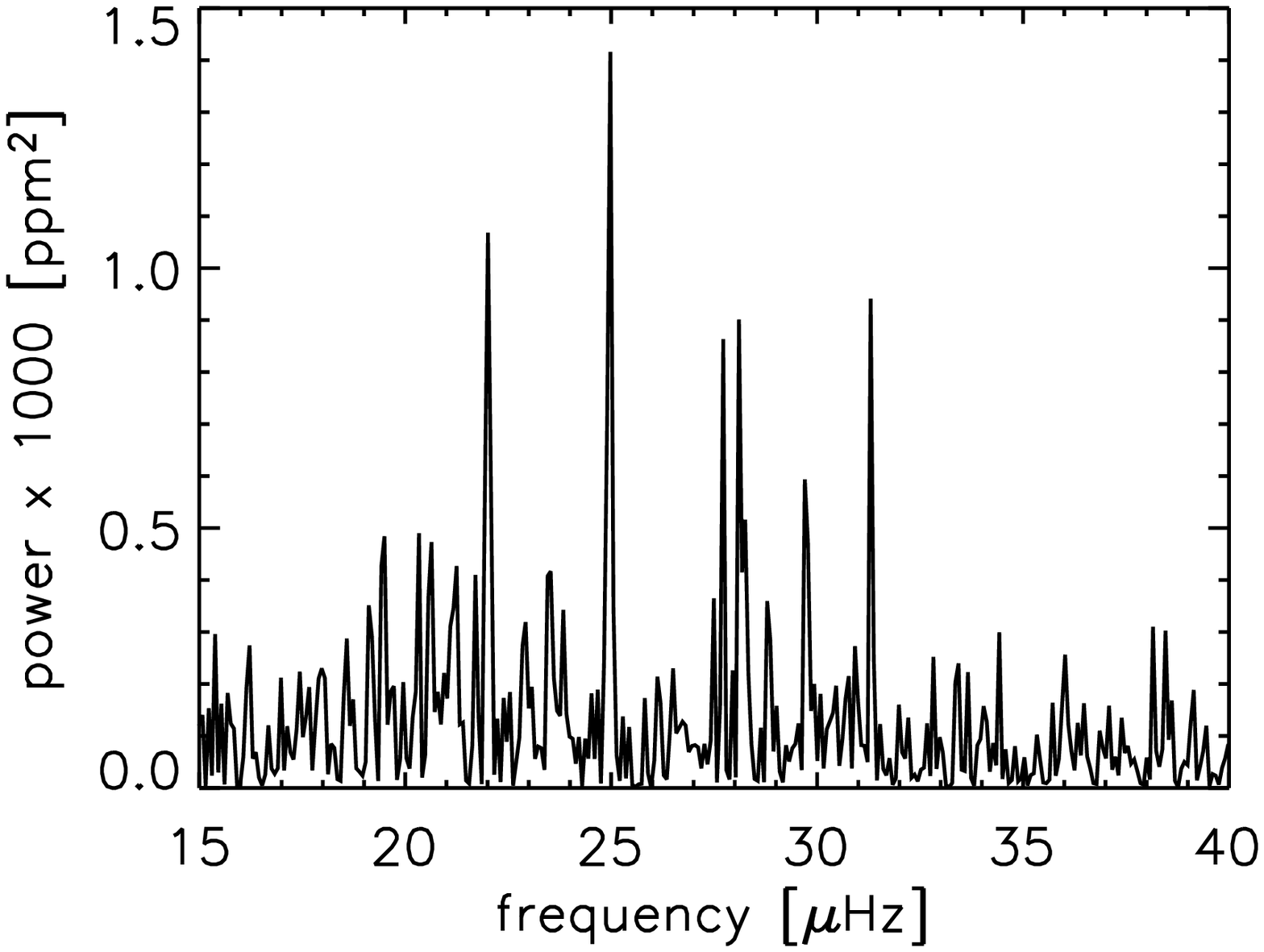}
\end{minipage}
\hfill
\begin{minipage}{\linewidth}
\centering
\includegraphics[width=\linewidth]{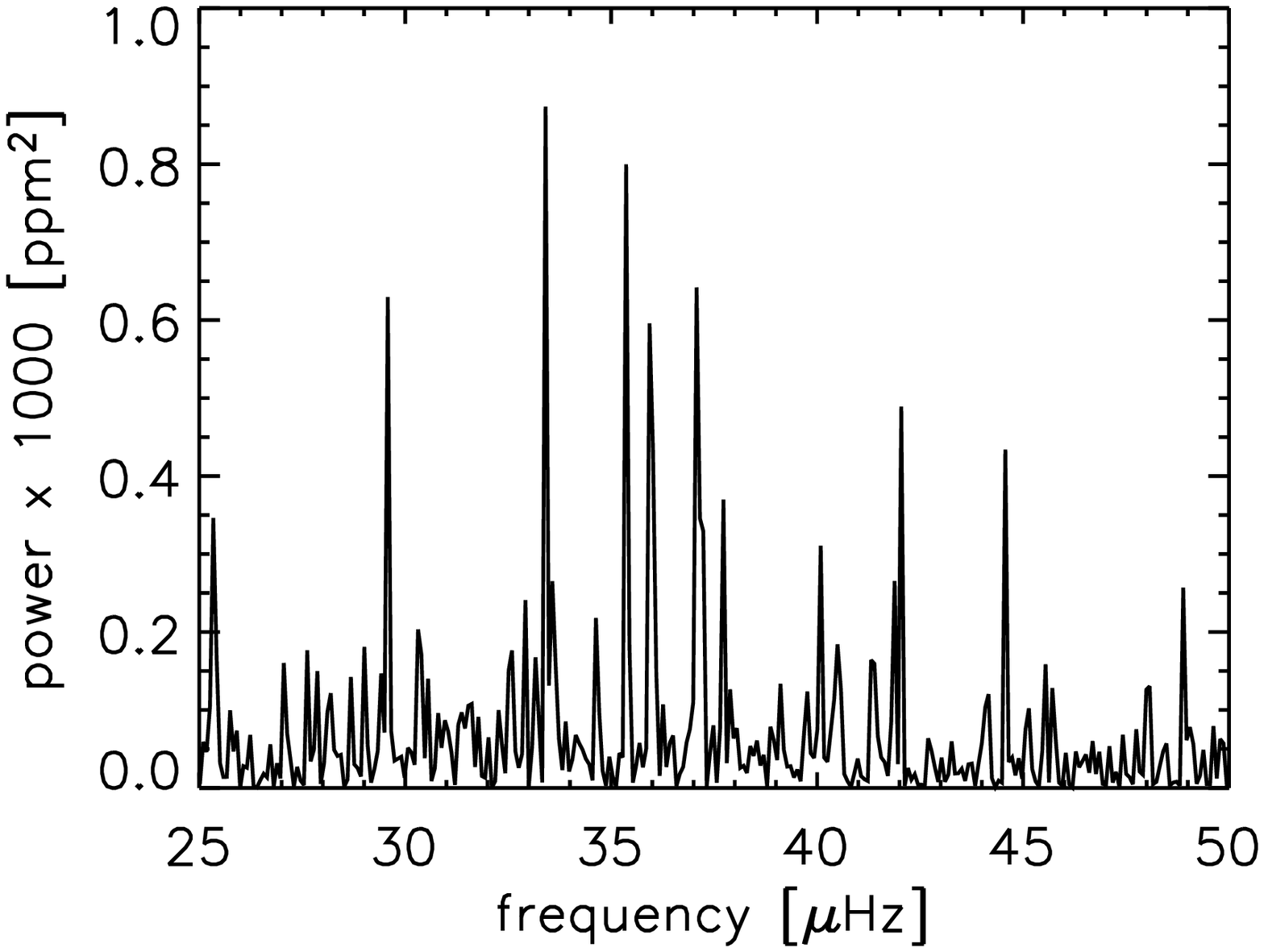}
\end{minipage}
\caption{Power spectrum of the CoRoT stars 100830101 (left) and 100597609 (right). A hint of regularity can be seen for 100830101, but the peaks around 28 $\mu$Hz could be due to two modes with long lifetime, or one mode with short lifetime. In the case of long lifetime modes, we can not identify both of them with the asymptotic relation, while the short lifetime interpretation would be very different from the long lifetimes of the narrow frequency peaks at 25 and 31 $\mu$Hz.}
\label{nonregular}
\end{figure}

\begin{figure}
\centering
\includegraphics[width=\linewidth]{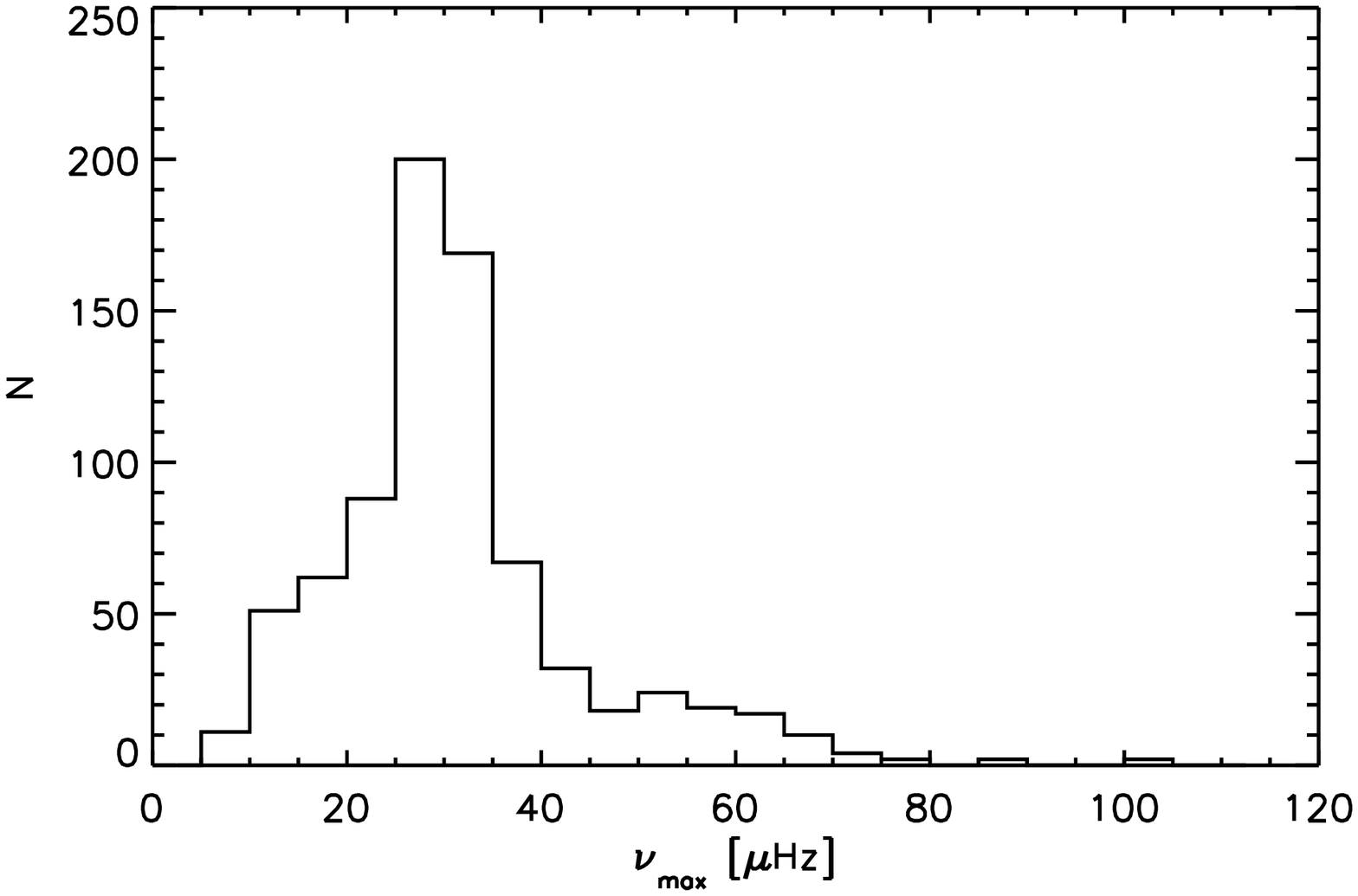}
\caption{Histogram of the frequencies of maximum oscillation power for the red-giant stars in the CoRoT LRc01 exoplanet field.}
\label{numax}
\end{figure}

The long (150 days) high precision photometric time series obtained with CoRoT provide evidence for non-radial modes with long lifetimes in red-giant stars, as discussed by \cite{deridder2009}. In Fig.~\ref{nonrad}, we show an example of the power spectrum and \'{e}chelle diagram of the CoRoT target 100483847. The narrow frequency peaks in the power spectrum indicate long mode lifetimes ($\geq$ 50 days). For high-order low-degree modes the radial modes of consecutive orders appear nearly equidistant, separated by the so-called large separation, with $\ell$ = 1 modes approximately half way in between the radial modes, and modes with $\ell$ = 2 close to the radial modes (small separation). Indeed, three vertical ridges are present in the \'echelle diagram, clearly indicating the presence of non-radial oscillations.

Not all red giants show these narrow regularly spaced frequency peaks, as can be seen in Fig.~\ref{nonregular}. For some stars at least a number of frequencies have a broader peak, indicating shorter mode lifetimes (of order a few days). Also in these cases, the large separation can not be determined accurately, if found at all.

For nearly 800 red giants, among the many stars observed during the first long run of CoRoT, solar-like oscillations have been detected \cite{hekker2009}. This allows for a statistical investigation of the characteristics of these solar-like oscillations for the first time. In particular, the frequency of maximum oscillation power ($\nu_{max}$) has been statistically studied because this parameter changes as a function of radius, which increases with stellar age, and can therefore probe stellar evolution. As shown by \cite{hekker2009}, the distribution of $\nu_{max}$ reveals that a large fraction of red giants have a $\nu_{max}$ in the range 20-40 $\mu$Hz, see also Fig.~\ref{numax}. 

\citet{hekker2009} provide an extensive discussion about possible observational biases and physical reasons why the $\nu_{max}$ distribution has a dominant peak between 20-40 $\mu$Hz. They arrive at the conclusion that this can be partly explained by granulation at  low frequencies and the decreasing height of the oscillations at high frequencies, but that the population density of the observed stars is also an important factor. This is confirmed by population studies performed by \cite{miglio2009}, see also these proceedings, which show that stars with 20~$\mu$Hz~$\leq$~$\nu_{max}$~$\leq$~40~$\mu$Hz are mainly He-burning red-clump stars.

\section{Prospects}
So far the CoRoT data have already greatly increased our knowledge of solar-like oscillations in red-giant stars, both from an observational as well as from a theoretical point of view. Apart from the results discussed here, detailed studies of individual stars, an investigation into mass and radius determination from oscillation properties \cite{kallinger2009} and a statistical investigation in mode lifetimes are under way. These observational efforts are accompanied by theoretical investigations  using adiabatic \cite{dupret2009} and non-adiabatic codes to simulate the stars, as well as stellar population studies \cite{miglio2009} . 

Notwithstanding these efforts, there still remain many open questions about the internal structures of red giants, and the different processes taking place in different layers and between layers in these stars. The excitation and damping of these oscillations, the time scales at which these process occur and their connection with convection and granulation are not well understood yet. As the sizes of the convective cells likely depend on the depth of the convection zone, it might also be very interesting to measure the depth of this zone. Furthermore, it would be very interesting to be able to infer whether the star is in the hydrogen-shell burning or helium-burning phase. With a detailed picture of the internal structure and mass of the red giants, it might also be possible to trace back their evolution and determine their primordial rotation velocity or infer the difference in internal structure due to different chemical composition.\newline
\newline
The future looks very promising concerning dedicated instrumentation for asteroseismology. The CoRoT satellite is still taking data and Kepler has recently started its observations. Both satellites observe many relatively faint stars (roughly 10 - 16 mag in V), which has the advantage of obtaining a statistically significant sample, but the disadvantage that stellar parameters, such as effective temperature, surface gravity and metallicity are less accurate, if known at all. 

For bright targets, for which it is relatively easy to obtain stellar parameters, the MOST satellite (Micro variability and Oscillations of STars) has been taking valuable time series data of a couple of weeks for several red giants. Furthermore, dedicated instrumentation is currently under development. BRITE-Constellation consists of two small satellites with different colour filters, which aim to take simultaneous two-colour photometry of bright targets (2-6 mag in V). Depending on the launch time, which is scheduled in 2010, there are fields that can be observed year round, while other fields can be observed for at least half a year. Another development is SONG (Stellar Oscillations Network Group), which aims for a dedicated network of 1~m class telescopes with high-resolution spectrographs for radial-velocity measurements. The prototype of these telescopes is currently being build at Tenerife, Spain. \newline
\newline
We believe that with MOST, CoRoT and Kepler running, and SONG and BRITE-Constellation on their way, significant progress in the field of asteroseismology, in particular with respect to red giants, can be expected.

%%%%%%%%%%%%%%%%%%%%%%%%%%%%%%%%%%%%%%%%%%%%%%%%
%% BACKMATTER
%%%%%%%%%%%%%%%%%%%%%%%%%%%%%%%%%%%%%%%%%%%%%%%%

\begin{theacknowledgments}
SH wants to thank M. Mooij for carefully reading the manuscript and valuable comments, which helped to improve the manuscript considerably.
 %TK and WWW acknowledges support by the Austrian Research Promotion Agency (FFG), and the Austria Science Fund (FWF P17580). FC is a postdoctoral fellows of the Fund for Scientific Research Flanders. APH acknowledges the support of grant 50OW0204 from the Deutsches Zentrum f\"ur Luft- und Raumfahrt e. V. (DLR). 
\end{theacknowledgments}

\bibliographystyle{aipproc}
\bibliography{HekkerSantaFebib}

\end{document}